\documentclass[12pt,a4paper]{article}
\usepackage{amssymb,amsfonts,amsmath,amsthm}
\usepackage{hyperref}
\author{Alexey~A.~Magazev\thanks{Omsk State Technical University, Omsk, Russia; e-mail: magazev@gmail.com}\ \ and Igor~V.~Shirokov\thanks{Omsk State Technical University, Omsk, Russia; e-mail: iv\_shirokov@mail.ru}}

\title{A method of integration for classical and quantum equations based on the connection between canonical transformations and irreducible representations of Lie groups}
\date{}

\newtheorem{lemma}{Lemma}
\newtheorem{theorem}{Theorem}

\begin{document}
\maketitle

\begin{abstract}
We propose a method for integrating the right-invariant geodesic flows on Lie groups based on the use of a special canonical transformation in the cotangent bundle of the group. We also describe an original method of constructing exact solutions for the Klein -- Gordon equation on unimodular Lie groups. Finally, we formulate a theorem which establishes a connection between the special canonical transformation and irreducible representations of Lie group. This connection allows us to consider the proposed methods of integrating for classical and quantum equations in the framework of a unified approach.
\end{abstract}

\section{Introduction}
Numerous researchers in the field of theoretical and mathematical physics often face the problem of integration of differential equations describing various classical and quantum systems. At the same time traditional approaches to solving the classical equations is essentially different from methods of integrating the quantum equations. For example, the general method of integrating the geodesic flows on pseudo-Riemannian manifolds is based on the symplectic reduction procedure \cite{Abr78}, while the problem of finding exact solutions for relativistic wave equations (Klein -- Gordon equation, Dirac equation, etc.) is commonly resolved with the technique of separation of variables \cite{Mil77, BagGit90}.

In this work we discuss the problem of integrating classical and quantum equations on manifolds of Lie groups. Namely, we shall consider Hamilton's equations for the right-invariant geodesic flows on Lie groups and the corresponding Klein -- Gordon equation.

The integrable geodesic flows on Lie groups were investigated by many prominent mathematicians such as V.~Arnold \cite{Arn66}, S.~Manakov \cite{Man76}, A.~Mishenko and A.~Fomenko \cite{MisFom78} and others. Here we describe a constructive method for integration of right-invariant geodesic flows in quadratures based on the ideas from the work \cite{MagShi03}. Hereafter, following the noncommutative integration method of linear differential equations \cite{ShaShi95}, we develop a technique for constructing the general solution of the Klein -- Gordon equation on unimodular Lie groups. In conclusion, we establish a profound connection between this methods which allows us to solve the problems of integrating the classical and quantum equations on Lie group in the framework of a unified approach.

%%%%%%%%%%%%%%%%%%%%%%%%%%%%%%%%%%%%%%
%
\section{Symmetries of classical and quantum equations on pse\-udo-Riemannian manifolds}\label{MagShi:sec-02}
%
%%%%%%%%%%%%%%%%%%%%%%%%%%%%%%%%%%%%%%
Let $(M,g)$ be a pseudo-Riemannian $n$-dimensional  manifold. We denote by $U \subset M$ a coordinate chart that trivializes the cotangent bundle $T^* M$, i.e. $T^* M|_U \simeq U \times \mathbb{R}^n$; the corresponding local coordinates are labelled as $(x^1, \dots, x^n, p_1, \dots, p_n)$. 

The \textit{geodesic flow} on $(M,g)$ is defined by the Hamilton's equations
\begin{equation}\label{MagShi:eq-01}
\dot{x}^i = \frac{\partial H^{cl}}{\partial p_i},\quad
\dot{p}_i = - \frac{\partial H^{cl}}{\partial x^i},
\end{equation}
with the Hamiltonian $H^{cl}(x,p) = \frac12\, g^{ij}(x) p_i p_j$ (we assume Einstein summation convention). Here $g^{ij}$ is the inverse of the metric tensor $g_{ij}$: $g^{ik} g_{kj} = \delta^i_j$. 
%If $(M,g)$ is a four-dimensional Lorentzian manifold then Hamiltonian system (\ref{MagShi:eq-01}) describes the motion of a free particle in external gravitational field associated with the metric~$g$.

The simplest quantum analogue of Hamiltonian system (\ref{MagShi:eq-01}) is the \textit{Klein--Gordon equation}
\begin{equation}\label{MagShi:eq-02}
H \psi := \left (g^{ij} \nabla_i \nabla_j + m^2 + \zeta R \right ) \psi = 0.
\end{equation}
Here $\nabla_i$ is the covariant derivative in the direction of the coordinate vector field $\partial_i:=\partial/\partial x^i$, $m$ is a positive real parameter, $R$ is the scalar curvature of the pseudo-Riemannian manifolds~$(M,g)$. Parameter $\zeta$ is a dimensionless coupling constant. 
%In case of space--time metric Eq. (\ref{MagShi:eq-02}) describes the dynamics of massive spinless particles of the scalar field $\psi$ interacting with the external gravitational field.

Let $G$ be the group of motions of the pseudo-Riemannian manifold $(M,g)$. The assosiated Lie algebra $\mathfrak{g}$ is generated by \textit{Killing vectors} $\xi_a = \xi_a^i(x) \partial_i$ with the commutation relations
\begin{equation}
[\xi_a, \xi_b] = C_{ab}^c\, \xi_c.
\end{equation}
Here $C_{ab}^c$ are the \textit{structure constants} of the Lie algebra~$\mathfrak{g}$.

The Killing vectors $\xi_a$ allows us to construct constants of motion for Hamiltonian system~(\ref{MagShi:eq-01}) that are linear on the fibers of the cotangent bundle $T^* M$: 
\begin{equation}\label{MagShi:eq-03}
\xi_a^{cl}(x,p) = \xi_a^i(x) p_i.
\end{equation}
We shall call these functions \textit{Killing constants of motion}. Clearly the span of the Killing constants of motion forms a Lie algebra with respect to the canonical Poisson bracket $\{\cdot,\cdot\}$; this Lie algebra is isomorphic to the Lie algebra $\mathfrak{g}$:
\begin{equation}
\{ \xi^{cl}_a, \xi^{cl}_b \} = C_{ab}^c\, \xi^{cl}_c.
\end{equation}

Furthermore, the vector fields $\xi_a$ considered as differential operators in the functional space $C^{\infty}(M)$ are symmetry operators of Eq. (\ref{MagShi:eq-02}) since the equality
$$
[H, \xi_a] = 0,
$$
holds for all $a = 1, \dots, \dim \mathfrak{g}$.

Thus, two algebras are naturally associated with the group of motions of the pseudo-Riemannian manifold $(M,g)$. The first one is the Lie algebra of the Killing constants of motion defined by formula (\ref{MagShi:eq-03}); the second one is the symmetry algebra for the Klein--Gordon equation which is generated by the first-order differential operators $\xi_a$. Both of these algebras are isomorphic to the Lie algebra $\mathfrak{g}$ of the group $G$.

%%%%%%%%%%%%%%%%%%%%%%%%%%%%%%%%
\section{Simple transitive actions of Lie groups}\label{MagShi:sec-03}
%%%%%%%%%%%%%%%%%%%%%%%%%%%%%%%%

We assume that action of the group of motion $G$ on the pseudo-Riemannian manifold $(M,g)$ is \textit{simply transitive}. This means that for any $x_1, x_2 \in M$ there exists precisely one $a \in G$ such that $x_2 = x_1 a$ (we assume that the action of group $G$ is the right action). It follows that there is a smooth one-to-one correspondence between the points of $M$ and the elements of Lie group~$G$. Keeping this diffeomorphism in mind, we represent Eqs. (\ref{MagShi:eq-01}), (\ref{MagShi:eq-02}) in terms of invariant vector fields on the group $G$.

Let $e_1, \dots, e_n$ be a basis in the Lie algebra $\mathfrak{g}$. Denote by $\eta_a(x) := (R_{x^{-1}})_* e_a$ the right-invariant vector field on $G$ corresponding to the basis vector $e_a$. It is easy to prove that any right-invariant pseudo-Riemannian metric $g$ on $G$ can be defined by the condition
\begin{equation}
\label{MagShi:eq-04}
g(\eta_a, \eta_b) = \mathbf{G}_{ab},
\end{equation}
where $\mathbf{G} = (\mathbf{G}_{ab})$ is a constant symmetric non-degenerate $n$-by-$n$ matrix. In the general case only the left-invariant vector fields $\xi_a(x) = (L_x)_* e_a$ are the Killing vectors for this metric.

From Eq. (\ref{MagShi:eq-04}) it follows that the Hamiltonian of the geodesic flow for the right-invariant metric has the form
\begin{equation}
\label{MagShi:eq-11}
H^{cl}(x,p) = \frac{1}{2}\, \mathbf{G}^{ab} \eta_a^{cl}(x,p) \eta_b^{cl}(x,p),
\end{equation}
where $\eta_a^{cl}(x,p) := \eta_a^i(x) p_i$, $\mathbf{G}^{ab} \mathbf{G}_{bc} = \delta^a_c$. Since the function $H^{cl}$ is invariant with respect to the right action of $G$ on $T^* G$, we shall call the corresponding geodesic flow the \textit{right-invariant geodesic flow}.

It is easily shown that the Klein--Gordon equation for metric (\ref{MagShi:eq-04}) can be represent as
\begin{equation}
\label{MagShi:eq-14}
H \psi = \left [ \left ( \mathbf{G}^{ab} \eta_a + C_a \right ) \eta_b + m^2 + \zeta R \right ] \psi = 0,
\end{equation}
where $C_a := C_{ab}^b$. Note that the scalar curvature for a right-invariant metric on $G$ is a constant.

%%%%%%%%%%%%%%%%%%%%%%%%%%%%%%%%
\section{An integration method for the right-invariant geodesic flows}\label{MagShi:sec-04}
%%%%%%%%%%%%%%%%%%%%%%%%%%%%%%%%

In this section we discuss the problem of integration of arbitrary right-invariant geodesic flows on Lie groups in quadratures. In particular, we describe the method of reduction for the corresponding Hamiltonian system based on using the special canonical transformation in cotangent space $T^* G$.

Denote by $\mathfrak{g}^*$ the dual space to the Lie algebra $\mathfrak{g}$ of the group $G$. Let $e^1, \dots, e^n$ be the basis of $\mathfrak{g}^*$ that is the dual for basis $e_1, \dots, e_n$ of $\mathfrak{g}$: $e^i(e_j) = \delta^i_j$, $i,j = 1, \dots, n$. We define the \textit{right momentum mapping} $\mu_r: T^* G \rightarrow \mathfrak{g}^*$ by the formula
$$
\langle \mu_r(x,p), X \rangle := X^a \xi_a^{cl}(x,p),
$$
where $X = X^a e_a \in \mathfrak{g}$. This mapping satisfies the condition $\{ \mu_r \circ \varphi, \mu_r \circ \psi \} = \mu_r \circ \{\varphi, \psi\}_{\mathfrak{g}}$ for all $\varphi, \psi \in C^{\infty}(\mathfrak{g}^*)$; here $\{\cdot,\cdot\}_{\mathfrak{g}}$ denotes the \textit{Lie -- Poisson bracket} on dual space $\mathfrak{g}^*$:
\begin{equation}
\label{MagShi:eq-05}
\{\varphi, \psi\}_{\mathfrak{g}}(f) = C_{ab}^c\, f_c \frac{\partial \varphi(f)}{\partial f_a} \frac{\partial \psi(f)}{\partial f_b},\quad
f = f_a e^a.
\end{equation}
Similarly, we define the \textit{left momentum mapping} $\mu_l: T^* G \rightarrow \mathfrak{g}^*$:
$$
\langle \mu_l(x,p),X \rangle := X^a \eta_a^{cl}(x,p),
$$
which satisfies the analogous condition: $\{ \mu_l\circ \varphi, \mu_l\circ \psi \} = \mu_l \circ \{\varphi, \psi\}_{\mathfrak{g}}$, $\varphi, \psi \in C^{\infty}(\mathfrak{g}^*)$. 
%Both of these mappings are polar to each other, so the pair $(\mu_r, \mu_l)$ forms a so-called \textit{dual pair}. 

It can be proved that the mappings $\mu_r$ and $\mu_l$ are equivariant with respect to the co-adjoint action $\mathrm{Ad}^*$
of the group $G$. Recall that the \textit{co-adjoint representation} is defined by
$$
\langle \mathrm{Ad}^*_x f, X \rangle := \langle f, \mathrm{Ad}_{x^{-1}} X \rangle,
$$
for all $x \in G$, $X \in \mathfrak{g}$ and $f \in \mathfrak{g}^*$. It can easily be checked that right and left momentum mappings are connected by the transformation
$$
\mu_l(x,p) = \mathrm{Ad}^*_x\, \mu_r(x,p).
$$

It is known that any Poisson manifold can be split into a collection of symplectic leaves. A. Kirillov showed that the symplectic leaves of the Lie -- Poisson bracket coincide with the \textit{co-adjoint orbits} of the group $G$ \cite{Kir76}. Thus, any co-adjoint orbit is a homogeneous symplectic manifold. 
%Particularly, all co-adjoint orbits have even dimensions.

Let ${\cal O}_{\lambda}$ be the co-adjoint orbit passing through the element $\lambda \in \mathfrak{g}^*$. We denote by $\omega_{\lambda}$ the symplectic 2-form on orbit ${\cal O}_{\lambda}$ that is defined by the restriction of bracket (\ref{MagShi:eq-05}) to the orbit ${\cal O}_{\lambda}$. The 2-form $\omega_{\lambda}$ is called the \textit{Kirillov--Konstant form}.

It follows from Darboux's theorem that there are local coordinates $q = (q^1,\dots,q^m)$ and $\pi = (\pi_1, \dots, \pi_m)$ on orbit ${\cal O}_{\lambda}$ such that
\begin{equation}
\omega_{\lambda} = d\pi_\alpha \wedge dq^\alpha,\quad
\alpha = 1, \dots, m = \frac{1}{2}\, \dim {\cal O}_{\lambda}.
\end{equation}
These coordinates are called \textit{canonical coordinates}. Denote by $f_a(q,\pi; \lambda)$ the functions which define the transition to the canonical coordinates on the orbit ${\cal O}_{\lambda}$. We restrict attention now to the case of the $\pi$-linear transition to canonical coordinates
\begin{equation}
\label{MagShi:eq-07}
f_a(q,\pi; \lambda) = X_a^\alpha(q) \pi_\alpha + \chi_a(q;\lambda).
\end{equation}
The existence of such transition is possible if and only if there exists a subalgebra $\mathfrak{p} \subset \mathfrak{g}^\mathbb{C}$ such that (see \cite{Shi00}):
\begin{equation}
\label{MagShi:eq-06}
\dim \mathfrak{p} = \dim \mathfrak{g} - \frac{1}{2}\, \dim {\cal O}_\lambda,\quad
\langle \lambda, [\mathfrak{p},\mathfrak{p}] \rangle = 0.
\end{equation}
A subalgebra $\mathfrak{p}$ satisfying the conditions (\ref{MagShi:eq-06}) is called a \textit{polarization} of element $\lambda \in \mathfrak{g}^*$. Note that polarizations of regular elements in $\mathfrak{g}^*$ always exist \cite{Dix77}.

It was shown in \cite{Shi00} that the coordinates $q$ in formula (\ref{MagShi:eq-07}) have the following interpretation. Let $P$ be the connected Lie group, whose Lie algebra is $\mathfrak{p}$. Then $q = (q^\alpha)$ are the local coordinates on the left coset $Q = G/P$. In case of real $P$, the manifold $Q$ is a Lagrangian submanifold of the symplectic manifold ${\cal O}_{\lambda}$. The action of $G$ on the homogeneous space $Q$ is defined by the co-adjoint action the group on orbit ${\cal O}_{\lambda}$:
\begin{equation}
\label{MagShi:eq-08}
\mathrm{Ad}^*_x	f(q,\pi; \lambda) = f(q',\pi'; \lambda) \Rightarrow q' = x q.
\end{equation}
%Here $\varphi: G \times Q \rightarrow Q $ is a map that defines the action $G$ on $Q$.

Now we introduce the function
\begin{equation}
\label{MagShi:eq-22}
S^\lambda(x; q, \pi') = \int f_a(x q,\pi'; \lambda) \sigma^a(x),
\end{equation}
where $f_a(q,\pi;\lambda)$ are defined by (\ref{MagShi:eq-07}) and $\sigma^a(x)$ are the right-invariant 1-forms on Lie group such that $\sigma^a(\eta_b) = \delta^a_b$, $a,b = 1, \dots, \dim \mathfrak{g}$. The function (\ref{MagShi:eq-22}) is well-defined because the 1-form under the integral is closed.

\begin{lemma}
The function $S^\lambda(x; q,\pi')$ satisfies the following relations
\begin{equation}
\label{MagShi:eq-09}
\pi_\alpha = \frac{\partial S^\lambda(x; q,\pi')}{\partial q^\alpha},\quad
q'^\alpha = - \frac{\partial S^\lambda(x; q,\pi')}{\partial \pi'_\alpha},
\end{equation}
where the canonical variables $(q^\alpha,\pi_\alpha)$ and $(q'^\alpha, \pi'_\alpha)$ are connected by transformation (\ref{MagShi:eq-08}).
\end{lemma}

Let $J = (J_1, \dots, J_r)$, $r = \mathrm{codim}\,{\cal O}_\lambda$, be local coordinates in an open neighborhood of the orbit space $\mathfrak{g}^* / G$. Denote by $\lambda(J)$ some local section of the bundle $\mathfrak{g}^* \rightarrow \mathfrak{g}^*/G$. We shall consider the function $S^{\lambda(J)}(x; q,\pi')$ as a \textit{generating function} of canonical transformation in the cotangent bundle $T^* G$. This canonical transformation is implicitly defined by Eqs. (\ref{MagShi:eq-09}) and the additional equalities
\begin{equation}
\label{MagShi:eq-10}
p_i = \frac{\partial S^{\lambda(J)}(x; q,\pi')}{\partial x^i},\quad
\tau^\mu = \frac{\partial S^{\lambda(J)}(x; q,\pi')}{\partial J_\mu},
\end{equation}
where $\mu = 1, \dots, r$. Thus, we have the smooth one-to-one coordinate transformation $(x,p) \leftrightarrow (q,\pi,q',\pi',J,\tau)$ which preserves the canonical symplectic form on $T^* G$:
$$
dp_i \wedge dx^i = d\pi_\alpha \wedge dq^\alpha + d\pi'_\alpha \wedge dq'^\alpha + dJ_\mu \wedge d\tau^\mu.
$$

After the canonical transformation (\ref{MagShi:eq-09}), (\ref{MagShi:eq-10}) the Hamiltonian (\ref{MagShi:eq-11}) of the right-invariant geodesic flow is converted to the function $\tilde{H}^{cl}(q',\pi';J) = \mathbf{G}^{ab} f_a(q',\pi'; \lambda(J)) f_b(q',\pi'; \lambda(J))/2$. The correspondence Hamiltonian system has the form
\begin{equation}
\label{MagShi:eq-12}
\dot{q}'^\alpha = \frac{\partial \tilde{H}^{cl}(q',\pi';J)}{\partial \pi'_\alpha},\quad
\dot{\pi}'_\alpha = \frac{\partial \tilde{H}^{cl}(q',\pi';J)}{\partial q'^\alpha},
\end{equation}
\begin{equation}
\label{MagShi:eq-13}
\dot{q}^\alpha = \dot{\pi}_\alpha = \dot{J}_\mu = 0,\quad
\dot{\tau}^\mu = \frac{\partial \tilde{H}^{cl}(q',\pi';J)}{\partial J_\mu}.
\end{equation}
Clearly, the integrability of system (\ref{MagShi:eq-12}), (\ref{MagShi:eq-13}) is equivalent to the integrability of its subsystem (\ref{MagShi:eq-12}).

Recall that an \textit{index} $\mathrm{ind}\,\mathfrak{g}$ of the Lie algebra $\mathfrak{g}$ is defined as the codimension of a regular co-adjoint orbit in $\mathfrak{g}^*$. The next theorem gives the integrability criterion for an arbitrary right-invariant geodesic flow on the Lie group $G$.
\begin{theorem}
An arbitrary right-invariant geodesic flow on $T^* G$ is integrable in quadratures if and only if
\begin{equation}
\label{MagShi:eq-20}
\frac{1}{2} \left ( \dim \mathfrak{g} - \mathrm{ind}\,\mathfrak{g} \right ) \leq 1.
\end{equation}
\end{theorem}

%%%%%%%%%%%%%%%%%%%%%%%%%%%%%%%%%%
\section{Integration of the Klein -- Gordon equation on Lie groups}\label{MagShi:sec-05}
%%%%%%%%%%%%%%%%%%%%%%%%%%%%%%%%%%

Now we consider the problem of integration for the Klein -- Gordon equation on unimodular Lie groups. The most efficient way for constructing exact solutions of the Klein -- Gordon equation is the \textit{noncommutative integration method} of linear differential equations suggested by A. Shapovalov and I. Shirokov~\cite{ShaShi95}. In contrast with the well-known method of separation of variables the noncommutative integration method uses the first-order symmetry algebra of the Klein--Gordon equation to the maximal extent possible.

The basic element of the noncommutative integration method is the so-called $\lambda$--representation of Lie algebra~$\mathfrak{g}$. We describe the procedure of the construction of this representation.

Let $\mathfrak{p}$ be a polarization of the regular element $\lambda \in \mathfrak{g}^*$, and let $\{e_A\}$ be a basis in $\mathfrak{p}$; $A = 1, \dots, \dim \mathfrak{p}$. We define the functional subspace $L(G,\mathfrak{p})$ of solutions of the system
$$
( \eta_A(x) - i \lambda_A) \psi(x) = 0,\quad \psi \in C^\infty(G).
$$
There is a local isomorphism $\iota: L(G, \mathfrak{p}) \rightarrow L(Q,\mathfrak{p})$ where $L(Q,\mathfrak{p})$ is the space of complex-valued functions on the mixed manifolds $Q$ (see \cite{Kir76}). As far as the functional subspace $L(G,\mathfrak{p})$ is invariant with respect to the right regular representation of Lie group $G$, the linear operators
\begin{equation}
\label{MagShi:eq-15}
\ell_a := \iota \circ \xi_a \circ \iota^{-1}
\end{equation}
are well-defined. The operators (\ref{MagShi:eq-15}) realize an irreducible representation of the Lie algebra $\mathfrak{g}$ which is called the $\lambda$--\textit{representation}. Note that in local coordinates on the manifold $Q$ the operators $\ell_a$ are ihhomogeneous first-order operators depending on $\dim Q = (\dim \mathfrak{g} - \mathrm{ind}\,\mathfrak{g})/2$ independent variables $q = (q^\alpha)$.

Let $T^\lambda$ be the lift of the $\lambda$--representation to a local representation of the Lie group $G$:
$$
\frac{d}{dt}\,( T^\lambda_{\exp(t e_a)}\varphi)(q) \big|_{t = 0} = \ell_a(q; \lambda).
$$
By definition
$$
\left( T^\lambda_x \varphi \right )(q) = \int \limits_Q {\cal D}^\lambda_{qq'}(x) \varphi(q') d\mu(q'),
$$
where ${\cal D}^\lambda_{qq'}(x)$ are the \textit{matrix elements} of the representation $T^\lambda$. Here $d\mu(q)$ is a measure on manifold $Q$. We choose the measure $d\mu(q)$ in such a way that the $\lambda$--representation operators are anti-Hermitian: $\ell_a^{\dagger} = - \ell_a$. In this case the representation $T^\lambda$ is a unitary irreducible representation of Lie group $G$.

It is easy to verify that the matrix elements of representation $T^\lambda$ satisfy the next system of equations
\begin{eqnarray}
\left ( \xi_a(x) + \ell_a(q; \lambda) \right ) {\cal D}^\lambda_{qq'}(x) = 0,\\
\left ( \eta_a(x) + \overline{\ell_a(q'; \lambda)} \right ) {\cal D}^\lambda_{qq'}(x) = 0.
\end{eqnarray}
Moreover, the matrix elements ${\cal D}^\lambda_{qq'}(x)$ obey the following completeness and orthogonality conditions:
\begin{equation}
\label{MagShi:eq-16}
\int {\cal D}^\lambda_{qq'}(x)\, \overline{{\cal D}^{\tilde{\lambda}}_{\tilde{q}\tilde{q}'}(x)}\, d\mu(x) = \delta(q,\tilde{q}) \delta(q',\tilde{q}') \delta(\lambda,\tilde{\lambda}),
\end{equation}
\begin{equation}
\label{MagShi:eq-17}
\int {\cal D}^\lambda_{qq'}(x) \overline{{\cal D}^\lambda_{qq'}(\tilde{x})}\, d\mu(q) d\mu(q') d\mu(\lambda) = \delta(x,\tilde{x}).
\end{equation}
Here $d\mu(x)$ is the invariant Haar measure on group $G$, and $d\mu(\lambda)$ is the spectral measure of the Casimir operators of the $\lambda$--representation. By $\delta(x,\tilde{x})$, $\delta(q,\tilde{q})$ and $\delta(\lambda,\tilde{\lambda})$ we denote the delta-functions for the measures $d\mu(x)$, $d\mu(q)$ and $d\mu(\lambda)$ respectively.

The relations (\ref{MagShi:eq-16}) and (\ref{MagShi:eq-17}) allows us to define the direct and inverse Fourier transformations as follows:
\begin{equation}
\label{MagShi:eq-18}
\hat{\psi}_\lambda(q,q') = \int {\cal D}^\lambda_{qq'}(x) \psi(x) d\mu(x),
\end{equation}
\begin{equation}
\label{MagShi:eq-19}
\psi(x) = \int \overline{{\cal D}^\lambda_{qq'}(x)} \hat{\psi}_\lambda(q,q') d\mu(q) d\mu(q') d\mu(\lambda). 
\end{equation}
Note that the actions of the operators $\xi_a$ and $\eta_a$ after transformation (\ref{MagShi:eq-18}) are mapping to actions of the corresponding $\lambda$--representation operators
\begin{eqnarray*}
\xi_a(x) \psi(x)\ \leftrightarrow & \ell_a(q; \lambda) \hat{\psi}_\lambda(q,q'),\\
\eta_a(x) \psi(x)\ \leftrightarrow & \overline{\ell_a(q'; \lambda)} \hat{\psi}_\lambda(q,q').
\end{eqnarray*}

Let $\psi(x)$ be an arbitrary solution of the Klein -- Gordon equation (\ref{MagShi:eq-14}). Using the decompositions (\ref{MagShi:eq-18}) and (\ref{MagShi:eq-19}) we obtain the equation for an unknown function $\hat{\psi}_\lambda(q,q')$:
$$
\left [ \mathbf{G}^{ab} \ell_a(q'; \lambda) \ell_b(q'; \lambda) + m^2 + \zeta R \right ] \hat{\psi}_\lambda(q,q') = 0.
$$
Note that the variables $q = (q^\alpha)$ enter to this equation as parameters.

We say that Eq. (\ref{MagShi:eq-14}) is \textit{integrable} if the problem of finding its general solution is reduced to calculation of quadratures and to solving ODE's. Keeping in mind this definition, we get the next result.

\begin{theorem}
The Klein -- Gordon equation (\ref{MagShi:eq-14}) on the unimodular Lie group $G$ is integrable with respect to an arbitrary right-invariant metric if and only if the inequation (\ref{MagShi:eq-20}) is satisfied.
\end{theorem}

%%%%%%%%%%%%%%%%%%%%%%%%%%%%%%%%%%%%%
\section{The connection between the generating function and the matrix elements of $\lambda$--representation}\label{MagShi:sec-06}
%%%%%%%%%%%%%%%%%%%%%%%%%%%%%%%%%%%%%

We define a measure $d\mu(\pi)$ by the condition
$$
\int e^{i (q - q')\pi} d\mu(\pi) = \delta(q,q').
$$
The following theorem gives the relation between the generating function $S^{\lambda}(x; q, \pi')$ of the special canonical transformation in $T^* G$ and the matrix elements ${\cal D}^\lambda_{qq'}(x)$ of the irreducible representation $T^\lambda$.

\begin{theorem}
\label{MagShi:th-01}
The matrix elements of the irreducible representation $T^\lambda$ of the group $G$ can be expressed as:
\begin{equation}
\label{MagShi:eq-21}
{\cal D}^\lambda_{qq'}(x) = \left| \frac{\partial (xq)}{\partial q} \right |^{-1/2} \int e^{i (q' \pi - S^\lambda(x; q, \pi) )} d\mu(\pi).
\end{equation} 
\end{theorem}

The theorem \ref{MagShi:th-01} is of fundamental importance in the solving of problem of integrating classical and quantum equations on Lie groups. Indeed, formula (\ref{MagShi:eq-21}) doesn't give only the rule for construction of the matrix elements ${\cal D}^\lambda_{qq'}(x)$, but also allows us to consider the methods of integration, proposed in sections \ref{MagShi:sec-04} and \ref{MagShi:sec-05},  in the framework of a unified approach. In particular, Theorem \ref{MagShi:th-01} has explained the fact that the integrability criterion for right-invariant geodesic flows on $G$ coincides with the integrability criterion for the corresponding Klein -- Gordon equations.

%%%%%%%%%%%%%%%%%%%%%%%%%%%%%%%%
\section{Conclusion}
%%%%%%%%%%%%%%%%%%%%%%%%%%%%%%%%

In this work we have suggested an integration method for the right-invariant geodesic flows on Lie groups which, in contrast to the symplectic reduction technique, is based on the use of the special canonical transformation in the cotangent bundle of group. 

We also have described an original method of constructing exact solutions for the Klein -- Gordon equation on unimodular Lie groups. According to this method, the procedure of integrating the Klein -- Gordon equation involves constructing of irreducible representations of Lie group and using the decomposition of the solution space into a sum of irreducible components. 

The main result of our work is Theorem \ref{MagShi:th-01} which establishes a fundamental connection between the proposed integration methods.
%%%%%%%%%%%%%%%%%%%%%%%%%%%%%%%%
\section*{Acknowledgement}
This research has been supported by Russian Foundation for Basic Research (grant 14-07-00272-a).

%%%%%%%%%%%%%%%%%%%%%%%%%%%%%%%%


\begin{thebibliography}{99}
\bibitem{Abr78} Abraham R. et al. 1978 {\it Foundations of mechanics} (Addison-Wesley Publishing Company, Inc.)

\bibitem{Mil77} Miller Jr, W. 1977 {\it Symmetry and separation of variables} (Addison-Wesley, Reading, Massachusetts)

\bibitem{BagGit90} Bagrov V.G. and Gitman D. 1990 {\it Exact solutions wave equations} (Springer) 

\bibitem{Arn66} Arnold V. I. 1966 {\it Annales de l'institut Fourier} {\bf 16}(1) 319--361

\bibitem{Man76} Manakov S. 1976 {\it Functional Analysis and Its Applications} {\bf 10}(4) 328--329

\bibitem{MisFom78} Mishchenko A. S. and Fomenko A. T. 1978 {\it Izvestiya Rossiiskoi Akademii Nauk. Seriya Matematicheskaya}. {\bf 42}(2) 396--415

\bibitem{MagShi03} Magazev A. A. and Shirokov I. V. {\it Theoretical and Mathematical Physics} {\bf 136}(3) 1212--1224

\bibitem{ShaShi95} Shapovalov A. V. and Shirokov I. V. {\it Theoretical and Mathematical Physics} {\bf 104}(2) 921--934

\bibitem{Kir76} Kirillov A. A. 1976 {\it Elements of the Theory of Representations} (Berlin: Springer-Verlag)

\bibitem{Shi00} Shirokov I. V. 2000  {\it Theoretical and Mathematical Physics} {\bf 123}(3) 754--767

\bibitem{Dix77} Dixmier J. 1977 {\it Enveloping algebras} (Newnes)

\end{thebibliography}
\end{document}